\documentstyle[11pt,newpasp,twoside,epsf]{article}
\begin{document}
\title{Cepheid and SNIa Distance Scales.}
\author{T. Shanks, P.D. Allen, F. Hoyle }
\affil{The Department of Physics, University of Durham, South Road, Durham DH1 3LE, England}
\author{N.R. Tanvir}
\affil{The Department of Physics \& Astronomy, Univ. of Hertfordshire}

\begin{abstract}
We first discuss why there remains continuing, strong motivation to
investigate Hubble's Constant. Then we review new evidence from an
investigation of the Galactic Open Clusters containing Cepheids by
Hoyle et al. that the metallicity dependence of the Cepheid P-L
relation is stronger than expected. This result is supported by a new
analysis of mainly HST Distance Scale Key Project data which shows a
correlation between host galaxy metallicity and the rms scatter around
the Cepheid P-L relation. If Cepheids do have a significant
metallicity dependence then an already existing scale error for
Tully-Fisher distances becomes worse and the distances of the Virgo
and Fornax clusters extend to more than 20Mpc, decreasing the value of
H$_0$. Finally, if the Cepheids have a metallicity dependence then so
do Type Ia Supernovae since the metallicity corrected Cepheid
distances to eight galaxies with SNIa now suggests that the SNIa peak
luminosity is fainter in metal poor galaxies. As well as having
important implications for H$_0$, this would also imply that the
evidence for a non-zero cosmological constant from the SNIa Hubble
Diagram may be subject to corrections for metallicity which are as big
as the effects of cosmology.
\end{abstract}

\section{Status  of Extragalactic Distance Scale}

One major motivation for studying Hubble's Constant is the complicated
nature of the current standard model in cosmology, $\Lambda$-CDM. In
this model, to order of magnitude, $\Omega_{baryon}\approx
\Omega_{CDM}\approx \Omega_{\Lambda}$ and this seems unnatural. The
coincidence between the CDM and Baryon densities worried a few when
CDM was first introduced (Peebles 1984, Shanks 1985).  The coincidence
between $\Omega_{\Lambda}$ and the others worried many more (eg
Dolgov, 1983, Peebles and Ratra, 1988 and Wetterich, 1988). These
fine-tuning problems of the standard model are compounded by the fact
that the inflation model on which the standard model sits, was partly
based on a fine-tuning argument, the flatness-problem; to begin by
eliminating one fine tuning problem only to end up with several gives
the appearance, at least, of circular reasoning!

Shanks (1985, 1991, 1999) noted that a simpler  model immediately became 
available if H$_0$ actually lay below 50 kms$^{-1}$ Mpc$^{-1}$. An inflationary
model with $\Omega_{baryon}$=1 is then better placed to escape the baryon
nucleosynthesis constraint. Simultaneously, the low value of H$_0$ means that the
X-ray gas in the Coma cluster increases towards the Coma virial mass and the
lifetime of an Einstein-de Sitter Universe extends to become compatible with the
ages of the oldest stars. Given the historical uncertainty there has been in the
value of H$_0$, this provides clear motivation for investigating the distance
scale route  to a better determination of Hubble's Constant.

\begin{figure}
\plotfiddle{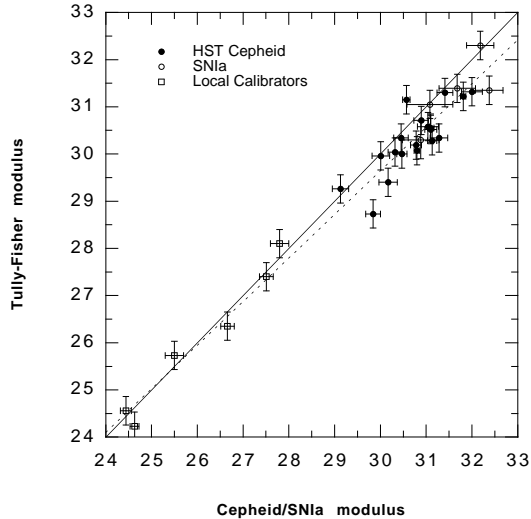}{3in}{0}{40}{40}{-120}{-50}
\caption{ A comparison between HST Cepheid and TF distances which suggests that
TF distances show a significant scale-error with the TF distance to galaxies at the 
distance of the Virgo cluster being underestimated by 22$\pm$5.2\%. The dashed line
shows the best fit with $(m-M)_{TF}= 0.915\pm0.036\times(m-M)_{Ceph}+2.204$.} 
\end{figure}

\section{A New Era for Determining H$_0$}

Some 25 galaxies have had Cepheids detected by HST. Seventeen of these were
observed by the HST Distance Scale Key Project (Freedman et al, 1994). Seven were
observed in galaxies with SNIa by Sandage and collaborators (eg Sandage et al,
1996) and M96 in the Leo I Group was observed by Tanvir et al (1995). In Fig. 1
we use these data to update the comparison of I-band TF distances of Pierce \&
Tully (1992) with HST Cepheid distances. As can be seen, the result implies that
TF distance moduli at Virgo underestimated by $\approx$25$\pm$5\%. This reduces
Tully-Fisher estimates of H$_0$ from $\approx$85 to
$\approx$65kms$^{-1}$Mpc$^{-1}$ (Giovanelli et al, 1997, Shanks 1997, Shanks,
1999, Sakai et al, 1999). The correlation of Cepheid residuals with line-width
suggests  TF distances may be Malmquist biased - possibly implying a bigger TF
scale error at larger distances. This clear problem for TF distances, which previously
has been the `gold standard' of secondary distance indicators, warns that errors
in the extragalactic distance scale may still be  seriously underestimated!

\section{NGC7790 Cepheid metallicity  dependence?}

New JKT 1.0m + CTIO 0.9m + UKIRT UBVK photometry of Cepheid Open Clusters by
Hoyle et al. (2001) has uncovered an anomaly in the NGC7790 UBV 2-colour
diagram, in that the F stars in the cluster show a strong  UV excess with respect
to zero-age main sequence stars (see Fig. 2). The result is confirmed by
independent photometric data (Fry, 1997, Fry and Carney, 1997) as shown in Fig. 7b
of Hoyle et al (2001). If the UV excess is caused by metallicity then NGC7790 would
have [Fe/H]$\approx$ -1.5 ! To keep the Galactic Cepheid P-L relation as tight
as previously observed implies that Cepheids may have a stronger metallicity
dependence, $\Delta M  \approx -0.66 \Delta[Fe/H]$, than previously expected, in
the sense that low metallicity Cepheids are intrinsically fainter. Currently we
are obtaining metallicities for the F stars in NGC7790 in order to confirm this
result.

\begin{figure}
\plotfiddle{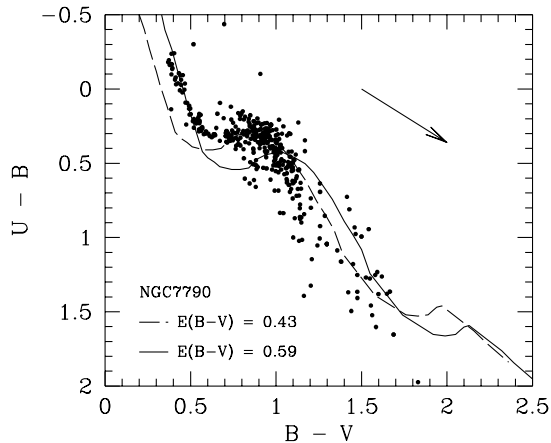}{3in}{0}{40}{40}{-120}{-50}
\caption{The UBV 2-colour plot from Hoyle, Shanks and Tanvir (2001) for the
Galactic open cluster, NGC7790, which contains 3 Cepheids. The data is not well
fitted by a solar metallicity, zero-age main sequence 2-colour diagram since
different values of the reddening, E(B-V) are implied by the B stars and the F
stars as shown. One interpretation is that the correct E(B-V) is given by the B
stars and that the F stars are showing UV-excess caused by the cluster
metallicity being significantly lower than Solar.}
\end{figure}

\section{HST Cepheid metallicity dependence}

Meanwhile, Allen \& Shanks (2001) have found an $\approx3\sigma$ correlation
between dispersion around the Cepheid P-L relation and galaxy metallicity for HST
Cepheid galaxies (see Fig. 3). This again suggests a strong Cepheid P-L
metallicity dependence and tends to support the results described above. The more
extended star-formation history of high metallicity galaxies may leave a wide
range of metallicities than in low metallicity galaxies like  the LMC, resulting
in a higher P-L dispersion if Cepheid luminosities at given period depend
strongly on metallicity.

Allen \& Shanks (2001) also obtain Cepheid distances  via truncated maximum
likelihood P-L fits to account for magnitude incompleteness caused by the
non-negligible scatter in the HST P-L relations. They found that Cepheid galaxy
distances at the limit of HST reach are too low. The higher than expected  P-L
dispersion  for distant, metal-rich galaxies accentuates this effect. The
conclusion is that current HST Cepheid distance moduli may be underestimated  by
more than 0.5 mag at the redshift of Virgo and Fornax due to  both metallicity
and statistical incompleteness bias. The TF distances discussed above are then
underestimates by approximately 1 magnitude.

\begin{figure}
\plotfiddle{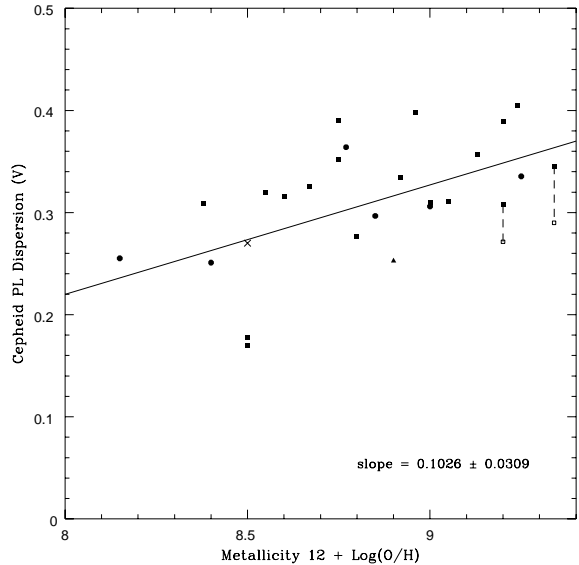}{3in}{0}{40}{40}{-120}{-50}
\caption{Relationship between mean r.m.s. dispersion (V-band) about the Cepheid
P-L relation and metallicity of HII regions in the vicinity of the Cepheids. Data
from the H$_{0}$ Key Project team is shown as squares. Sandage et al data is
shown as circles and the Tanvir et al data as a triangle. Finally, the LMC is
shown as a cross. Dashed vertical lines show the effect of removing outliers in
the P-L relations of two galaxies. The result of a least squares fit to the data
is also shown.}
\end{figure}

Eight HST Cepheid galaxies also have Type Ia distances. Correcting the Cepheid
scale for metallicity and incompleteness bias as above and then using these
distances to derive peak luminosities using the SNIa data from Gibson et al (2000),
implies a strong correlation between Type Ia peak luminosity and metallicity.
Such a scatter in SNIa luminosities could easily be disguised by magnitude
selection effects at moderate redshifts. At higher redshift the
correlation is in the right direction to explain away the need for a cosmological
constant in the Supernova Hubble Diagram results, since galaxies at high redshift
might be expected to have lower metallicity. Thus the conclusion is that if
Cepheids have strong metallicity dependence then so have SNIa and therefore SNIa
estimates of q$_0$ and H$_0$ may require  significant corrections for
metallicity.

\begin{figure}
\plotfiddle{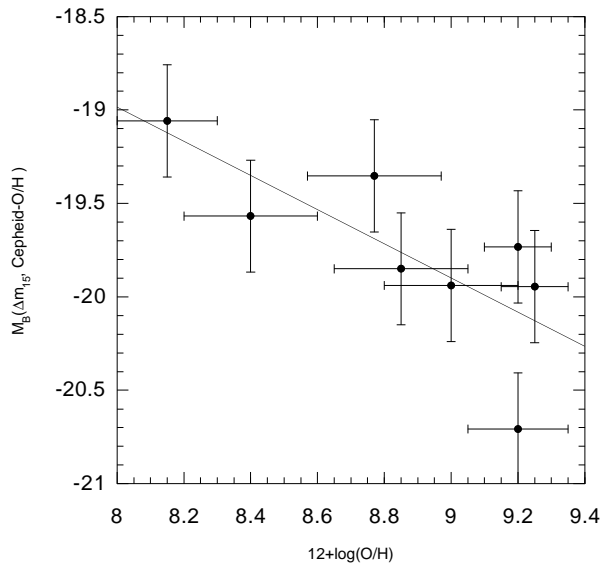}{3in}{0}{60}{60}{-180}{-200}
\caption{The SNIa absolute magnitude-metallicity relation using the SNIa
peak magnitudes of Gibson et al. (1999), now corrected for $\Delta m_{15}$  and
Cepheid metallicity. The least-squares fitted line,  $M_B=-0.92(\pm0.33)logW-11.65$, 
is also shown.} 
\end{figure}

\section{Conclusions - Implications for H$_0$ and SNIa}

Our conclusions are as follows:-

\begin{itemize}

\item Key Project  HST Cepheid distances imply Tully-Fisher distances at Virgo/Fornax are
underestimated by $\approx25\pm5$\%, reducing H$_0$ from $\approx$85 to
$\approx$65kms$^{-1}$Mpc$^{-1}$.

\item TF distances may be Malmquist biased,  suggesting there may be a 
bigger TF scale error at larger distances.

\item If the UV excess of F stars in open cluster NGC7790 is caused by low
metallicity then  Cepheids have  a strong metallicity dependence,  $\Delta M 
\approx -0.66  \Delta[Fe/H]$.

\item  Current HST Cepheid distances may be significantly underestimated at
Virgo/Fornax redshifts due to metallicity and magnitude incompleteness  bias, implying that
values of H$_0<$50kms$^{-1}$Mpc$^{-1}$  may still not be ruled out.

\item If Cepheids have a strong metallicity dependence then so have SNIa . Thus 
significant metallicity corrections may need to be applied to the Type Ia Hubble 
Diagram before reliable estimates of q$_0$ or H$_0$ can be made.

\end{itemize}

\acknowledgments We thank the HST Distance Scale Key Project for making the Cepheid data 
freely available.


\begin{references}

\reference Allen, P.D. \& Shanks, T. 2001, MNRAS, submitted.

\reference Dolgov, A.D., 1983, In `The Very Early Universe', Eds. Gibbons, G.W., Hawking, S., \& Siklos, S.T.C. CUP, pp. 449-458

\reference Freedman, W. L. et al., 1994, ApJ, 427, 628.

\reference Fry, A.M., 1997, PhD thesis, Univ. of North Carolina.

\reference Fry, A.M. \& Carney, B.W., 1997, AJ, 113, 1073. 

\reference Gibson, B. K. et al, 1999, ApJ, 512, 48.

\reference Giovanelli, R., Haynes, M.P., da Costa, L.N., Freudling, W., Salzer, J.J.,
       Wegner, G., 1997, ApJ, 477, L1.

\reference Hoyle, F., Shanks, T. \& Tanvir, N.R., 2001 MNRAS submitted, astro-ph/0002521.

\reference Peebles, P.J.E., 1984, ApJ, 284, 439.

\reference Peebles, P.J.E. \& Ratra, B. 1988, ApJ, 325, L17.

\reference Pierce, M.J. and Tully, R.B., 1992 ApJ, 387, 47.

\reference Sakai, S. et al, 1999, ApJ, 523, 540.

\reference Sandage, A.R., Saha, A., Tammann, G. A., Labhardt, L.,  Panagia, N., Macchetto, F. D.,
           ApJ, 1996, 460, L15.

\reference Shanks, T., 1985, Vistas in Astronomy, 28, 595.

\reference Shanks, T. et al., 1991, In `Observational Tests of Cosmological
Inflation', Eds. Shanks, T., Banday, A.J., Ellis, R.S., Frenk, C.S. \& Wolfendale,
A.W. pp. 205-210 Dordrecht:Kluwer.

\reference Shanks, T., 1997, MNRAS, 290, L77.

\reference Shanks, T., 1999, In Harmonising Cosmic Distance Scales, eds Egret D. 
\& Heck, A., PASP, pp. 230-234.

\reference Tanvir, N. R., Shanks, T., Ferguson, H.C. \& Robinson, D.R.T. 1995, Nature, 377, 27.


\reference Wetterich, C., 1988, Nucl. Phys. B302, 645.



\end{references}
\end{document}